\documentclass{ws-procs975x65} 
\begin{document}
\def \bea{\begin{eqnarray}}
\def \beq{\begin{equation}}
\def \eeq{\end{equation}}
\def \esix{E$_{\rm 6}$}
\def \ok{\overline{K}^0}
\def \s{\sqrt{2}}
\def \st{\sqrt{3}}
\def \sx{\sqrt{6}}

\title{Sterile Neutrinos in \esix}

\author{Jonathan L. Rosner$^*$}

\address{Enrico Fermi Institute, University of Chicago,\\
Chicago, Illinois 60637, USA\\
$^*$E-mail: rosner@hep.uchicago.edu\\
hep.uchicago.edu/\~{}rosner}

\begin{abstract}
The opportunity to accommodate three flavors of sterile neutrinos exists within
the exceptional group \esix.  Implications of this description are discussed.
\end{abstract}


\bodymatter

\section{Introduction \label{sec:intro}}

Sterile neutrinos are weak isosinglet neutrinos, visible through mixing
with one or more of the three ``active'' neutrinos $\nu_e,\nu_\mu,\nu_\tau$.
Several tentative indications exist that the three active neutrinos
aren't enough to fit all oscillation data; sterile neutrinos are one
possibility.  Present data prefer at least one sterile neutrino, but there are
tensions even with two.  In the grand unified group \esix~three sterile
neutrinos are natural; this talk explores some distinguishing features of such
a description.\cite{Rosner:2014cha}

We first review the shortcomings of a description with only three active
neutrinos (Sec.\ \ref{sec:ev}); this topic has been covered in greater detail
by C. Giunti.\cite{Giunti}  We then discuss mass matrices in \esix~and its
subgroups (Sec.\ \ref{sec:mm}), and their relevance for short-baseline
neutrino oscillation experiments (Sec.\ \ref{sec:sbl}).  One of the three
sterile neutrinos could play the role of a 7 keV dark matter candidate
(Sec.\ \ref{sec:dm}).  We conclude in Sec.\ \ref{sec:sum}.

\section{Evidence for sterile neutrinos \label{sec:ev}}

An early conflict with the picture of three active neutrinos was seen by the
LSND experiment at Los Alamos.\cite{Aguilar:2001ty}  The MiniBooNE experiment
at Fermilab confirmed this result (after a reanalysis of their data).%
\cite{AguilarArevalo:2007it,AguilarArevalo:2008rc,AguilarArevalo:2010wv,%
Aguilar-Arevalo:2013pmq}
The signal is mainly at low energy, falling below an initial energy cut of
475 MeV.  It is not clear whether the signal is $e^\pm$ or photons.  A possible
photon source would arise from a $Z$--$\omega$--$\gamma$ Wess-Zumino-Witten
(WZW) coupling\cite{WZ,W} giving rise to neutral-current coherent photon
production [Fig.\ 1(a)] off a nuclear target.\cite{Harvey:2007rd,Harvey:2007ca}
\begin{center}
\includegraphics[width=0.96\textwidth]{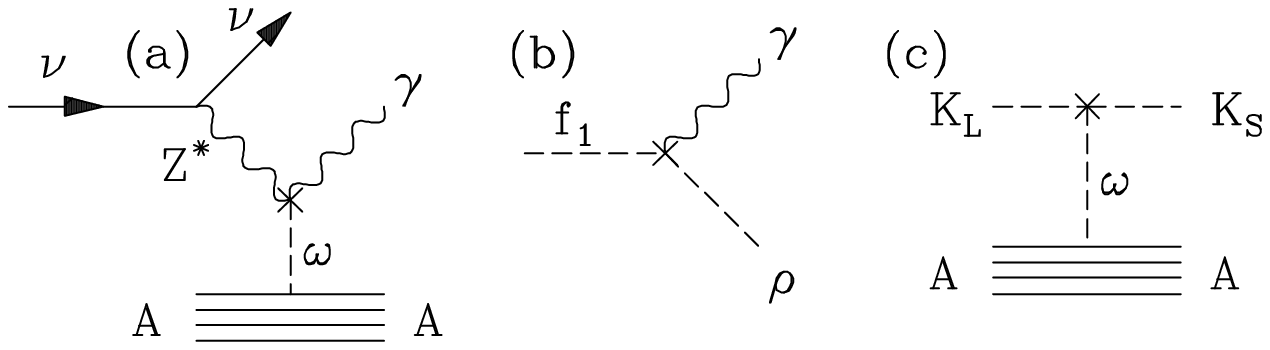}
\end{center}
A virtual $Z^*$ transforms as a $J^{PC} = 1^{++}$ $a_1$ meson.  The decays
$a_1^0(1260) \to \omega \gamma$ and the related decay $a_1^\pm(1260) \to
\rho^\pm \gamma$ are hard to look for.  Possible evidence for a related WZW
coupling comes from the decay $f_1(1285) \to \rho^0 \gamma$ [Fig.\ 1(b)];
the Mark III Collaboration at SPEAR observed this process\cite{Coffman:1989nk}
at a rate 12 times the non-WZW prediction.\cite{Babcock:1976hr,Rosner:1980ek}
A related term generates an $\omega K \bar K$ coupling giving rise to $K_S$
regeneration off a nuclear target [Fig.\ 1(c)].  One estimate of the
contribution of the process (a)\cite{Rosner:2015fwa} gives a rate about 1/4
that needed to explain the MiniBooNE result.

A claimed 6\% deficit with respect to expectations in the flux of reactor
neutrinos could be due to very-short-baseline neutrino oscillations.%
\cite{Mueller:2011nm,Mention:2011rk,Huber:2011wv}  A cautionary note%
\cite{Hayes:2013wra} identifies an additional uncertainty associated with
30\% of the flux coming from forbidden decays, whose intensity and energy
spectra are hard to evauate.

More evidence in favor of an anomaly comes from the use of $^{51}$Cr and
$^{37}$Ar radioactive sources to calibrate the SAGE and Gallex solar
neutrino detector,\cite{Abdurashitov:2009tn,Kaether:2010ag} finding an
observed/predicted ratio of $0.84\pm0.05$.\cite{Giunti:2012tn}
Both the reactor and the gallium anomalies could be due to short-baseline
neutrino oscillations with $\Delta m^2 = {\cal O}$(eV$^2$).  Such a large
splitting cannot be accommodated with three active neutrinos, whose masses
satisfy  $\Delta m^2_{21} \simeq 7.6 \times 10^{-5}$ eV$^2$; $|\Delta m^2_{32}|
\simeq 2.4 \times 10^{-3}$ eV$^2$.  A full set of constraints, including ones
from the BNL E-776, CDHSW, Daya Bay, ICARUS, KARMEN, MiniBooNE, MINOS, NOMAD,
OPERA, SciBooNE, and T2K experiments, is discussed in
Ref.\cite{Abazajian:2012ys} and by C. Giunti in this Conference.\cite{Giunti}
The absence of oscillations to one flavor ($N=1$) of sterile neutrino is
disfavored at the $6.3 \sigma$ level.

Even with more than one sterile neutrino a basic tension remains between
disappearance and appearance experiments.  In one fit\cite{Conrad:2012qt}
these are compatible only at the level of 0.008\%, mainly owing to a poor fit
to the low-energy MiniBooNE ``$e^\pm$'' signal.  For another fit, see
Ref.\cite{Kopp:2013vaa} An initial incompatibility between neutrino and
antineutrino fits, favoring $N > 1$, has been resolved with subsequent data, so there is no longer preference for more than one flavor of sterile
neutrino.\cite{Giunti}  Nevertheless, should such a need arise in the future,
the \esix~scheme provides a natural home for three sterile flavors ($N=3$).

\section{Mass matrices in \esix~and subgroups \label{sec:mm}}

The group SU(5)\cite{Georgi:1974sy} is the unique one of rank 4 containing
the standard model (SM) group SU(3)$_{\rm color} \times$ SU(2)$_{\rm L} \times$
U(1).  The quarks and leptons belong to $5^*+10$ representations; there is no
need for a right-handed neutrino.  The group SO(10)\cite{Fritzsch:1974nn}
contains SU(5); its 16-plet spinor contains the SU(5)
representations $5^*+10+1$, where the SU(5) singlet is a right-handed
neutrino $N$.  If this state is given a large Majorana mass, the corresponding
left-handed neutrino Majorana mass can be made very small (the {\it seesaw}
mechanism\cite{seesaw}).
The rank-5 nature of SO(10) implies the existence of an extra U(1) and the
possible observability of a $Z'$ at the TeV scale or above.

The exceptional group \esix\cite{Gursey:1975ki} contains SO(10).
Each fundamental 27-plet of \esix~contains 16 + 10 + 1 of SO(10).  The 10 of
SO(10) contains $5+5^*$ of SU(5), where the 5 contains a color-triplet weak
isosinglet quark and a color-singlet weak isodoublet lepton.  The singlet of
SO(10) (``$n$'') is a sterile neutrino candidate.  Since one needs three
27's to account for three families of ordinary quarks and leptons, there are
three sterile neutrinos in \esix.  The rank-6 nature of \esix~implies the
possibility of two extra U(1)'s or at least one linear combination surviving
symmetry breaking down to LHC energies.  The U(1) charges are defined as\\

\esix $\to$ SO(10) $\times$ U(1)$_\psi~(Q_\psi)$~;~~~
SO(10) $\to$ SU(5) $\times$ U(1)$_\chi~(Q_\chi)$~.\\

A $Z_\theta$ can couple to $Q_\psi \cos \theta + Q_\chi \sin \theta$.  The
combination $Q_N \equiv -(1/4) Q_\chi + (\sqrt{15}/4) Q_\psi$ vanishes for the
right-handed neutrino $N$.  A large Majorana $N$ mass is then permitted by
$Q_N$ conservation, enabling a seesaw mechanism with fermion
masses generated by a 27-plet Higgs representation.\cite{Ma:1995xk} The U(1)
charges for various members of a 27-plet are shown in Table 1.

\begin{table}
\tbl{U(1) charges of 27-plet fermions in \esix.}
{\begin{tabular}{c c c c} \toprule
   27 member   & $2\sqrt{6}~Q_\psi$ & $2\sqrt{10}~Q_\chi$ & $2\sqrt{10}~Q_N$ \\
(SO(10),SU(5)) &                   &                    &       \\ \colrule
$\nu_e(16,5^*)$ &        $-1$       &       ~~3          &       $-2$      \\
~~(16,10)      &         $-1$      &        $-1$        &       $-1$      \\
$N^c_e(16,1)$   &        $-1$       &       $-5$         &       ~~0       \\
$\nu_E(10,5^*)$ &        ~~2        &       $-2$         &       ~~3       \\
$N^c_E(10,5)$   &        ~~2        &       ~~2          &       ~~2       \\
    $n(1,1)$    &        $-4$       &       ~~0          &       $-5$      \\
\botrule
\end{tabular}}
\end{table}

The $Z_N$, coupling to $Q_N$, has characteristic branching fractions.  Within
a single family, 25\% of its decays are to ordinary fermions (above the
middle line in Table 2) while 75\% are to exotic fermions (below the middle
line).  These consist of a vector-like charged lepton $E$, its neutrino
$\nu_E$, an isosinglet quark $h$, and the sterile neutrino $n$.  If the
$Z_N$ is found at the LHC, it is a potential source of exotic quarks and
leptons.  The differences between left-handed (L) and right-handed (R)
couplings give rise to characteristic production and decay
asymmetries.\cite{Langacker:1984dc}

\begin{table}
\tbl{Branching fractions of $Z_N$ within a family.}
{\begin{tabular}{c r r r r} \toprule
 Decay  & \multicolumn{3}{c}{Helicity} & Percent \\
product &     L    &     R    &   Sum  & of total\\ \colrule
$e \bar e$ & 4/120 & 1/120 & 5/120 &  4.17 \\
$\nu_e \bar \nu_e$ & 4/120 & -- & 4/120 & 3.33 \\
$u \bar u$ & 3/120 & 3/120 & 6/120 &  5.00 \\
$d \bar d$ & 3/120 &12/120 &15/120 & 12.50 \\ \colrule
$E \bar E$ & 9/120 & 4/120 &13/120 & 10.83 \\
$\nu_E \bar \nu_E$ & 9/120 & 4/120 & 13/120 & 10.83 \\
$h \bar h$ &27/120 &12/120 &39/120 & 32.50 \\
$n \bar n$ &25/120 &   --  &25/120 & 20.83 \\
\botrule
\end{tabular}}
\end{table}

While $27 \times 27 = 27^* + 351 + 351'$, we wish to see what follows from
assuming $27^*$ dominates, which was a popular assumption in the early days of
string theory.\cite{Dine:1985vv,Breit:1985ud,Cecotti:1985by,Rosner:1985hx,%
Nandi:1985uh}  Some mass matrix elements will be absent as their $(Q_\psi,
Q_\chi)$ values aren't in $27^*$.  The U(1) charges for the product 27 $\times$
27 are shown in Table 3, where we have listed values of $(2\sqrt{6}~Q_\psi,
2\sqrt{10}~Q_\chi,2\sqrt{10}~Q_N)$. Blank entries denote charges not found in a
$27^*$-plet, implying a zero entry in the mass matrix.  The exception (in the
box) is a Majorana mass for the right-handed neutrino $N_e^c$, which must be
generated by a higher-dimension operator conserving $Q_N$.

\begin{table}
\tbl{U(1) charges (see text) in the product of two 27's of \esix.}
{\begin{tabular}{c|c c c c c} \toprule
 & $\nu_e$(--1,3,2) & $N^c_e$(--1,--5,0) & $\nu_E$(2,--2,3) & $N^c_E$(2,2,2)
 & $n$(--4,0,--5) \\ \colrule
$\nu_e$(--1,3,2) & -- & (--2,--2,--2) & -- & $(1,5,0)$ & -- \\
$N^c_e$(--1,--5,0) & (--2,--2,--2) & \framebox[0.85in]{(--,--,0)} & --
 & (1,--3,2) & -- \\
$\nu_E$(2,--2,3) & -- & -- & -- & (4,0,5) & (--2,--2,--2) \\
$N^c_E$(2,2,2) & (1,5,0) & (1,--3,2) & (4,0,5) & -- & (--2,2,--3) \\
$n$(--4,0,--5) & -- & -- & (--2,--2,--2) & (--2,2,--3) & -- \\ \botrule
\end{tabular}}
\end{table}

For simplicity we make two further assumptions.  First, we
let $\nu_E$ pair up with $N^c_E$ to obtain a large Dirac mass $M_{34}$.
Second, we assume an approximate $Z_2$ symmetry to suppress vacuum expectation
values stemming from SO(10) 16-plets in comparison with those from SO(10)
10's or singlets.  The mass matrix in the basis $(\nu_e,N^c_e,\nu_E,N^c_E,n)$,
where we have used small letters to denote entries with weak isospin $\Delta I
= 1/2$ and large letters to denote entries with $\Delta I = 0$, is
$$
{\cal M} = \left[ \begin{array}{c c c c c} 0 & m_{12} & 0 & M_{14} & 0 \cr
 m_{12} & M_{22} & 0 & m_{24} & 0 \cr 0 & 0 & 0 & M_{34}
 & m_{35} \cr
 M_{14} & m_{24}& M_{34} & 0 & m_{45} \cr 0 & 0 & m_{35} & m_{45} & 0
\end{array} \right]~.
$$
It is convenient to diagonalize this matrix with respect to the large
entry $M_{34}$, leading to
$$
{\cal M}'{=}\left[ \begin{array}{c c c c c}
 0 & m_{12} & M_{14}/\s & M_{14}/\s & 0 \cr
 m_{12} & M_{22} & m_{24}/\s & m_{24}/\s & 0 \cr
 M_{14}/\s & m_{24}/\s & M_{34} & 0 & (m_{35}{+}m_{45})/\s \cr
 M_{14}/\s & m_{24}/\s & 0 & -M_{34} & (m_{45}{-}m_{35})/\s \cr
 0 & 0 & (m_{35}{+}m_{45})/\s & (m_{45}{-}m_{35})/\s & 0
\end{array} \right]~.
$$
Now we can perturb about the three eigenvectors $[0,1,0,0,0]^T$,
$[0,0,1,0,0]^T$, and $[0,0,0,1,0]^T$ corresponding to the large eigenvalues
$M_{22},M_{34},{-}M_{34}$.  For the small masses, the resulting $2 \times 2$
mass matrix in the $(\nu_e,n)$ basis is
$$
{\cal S}_2 = \left[ \begin{array}{c c}
   -m_{12}^2/M_{22}    & -M_{14}m_{35}/M_{34} \cr
  -M_{14}m_{35}/M_{34} & -2m_{35} m_{45}/M_{34} \end{array} \right]~.
$$
We look for solutions with small mixing and $m_n > m_\nu$:
$$
\nu= \left[ \begin{array}{c} \cos\theta \cr \sin\theta \end{array} \right]~,~~
 n= \left[ \begin{array}{c} -\sin\theta \cr \cos\theta \end{array} \right]~,~~
t \equiv \tan \theta~,
$$
so we seek a small-$t$ solution of a quadratic equation in $t$, which in its
linearized form is
$$
t \simeq \left( \frac{m_{12}^2 M_{34}}{M_{14}m_{35}M_{22}}
 - \frac{2 m_{45}}{M_{14}} \right)^{-1}~.
$$
Barring accidental cancellations, after several steps we get $m_n >
m_\nu$ with small mixing if $M_{14} \ll m_{45}$ and
$$
\left | \frac{m_{35}m_{45}M_{22}}{M_{34}m_{12}^2} \right| > 1~,~~
\frac{m_{45}}{M_{14}} \gg 1~,
$$
The smallness of $M_{14}$ is curious but achievable via the approximate $Z_2$
symmetry mentioned earlier.

The neutrino mass matrix can be related to those for charged fermions at a
unification scale.  Thus, $m_{12}$ and $m_{35}$ are related to masses of quarks
of charge 2/3, while $m_{24}$, $m_{45}$, $M_{14}$, and $M_{34}$ are related to
charge --1/3 quark and charged lepton masses.  Specifically, for up-type
quarks, the U(1) charges of masses are (--2,--2,--2), corresponding to $m_{21}$
and $m_{35}$.  The relation of $m_{12}$ to $m_u$ is familiar from SO(10)
unification.  For down-type quarks and charged leptons, the correspondences are
(--2,2,--3)$\sim m_{45}$,~(1,5,0)$\sim M_{14}$,~(1,--3,2)$\sim m_{24}$,
and (4,0,5)$\sim M_{34}$.  In the absence of mixing, $m_{45}$ is related to
Dirac masses of $d$ and $e$, while $M_{34}$ is related to Dirac masses of
quarks and charged leptons in the 10 of SO(10).  Weak universality suggests
$|m_{24}|\ll |m_{45}|$ (because isosinglet impurities in left-handed charged
leptons and down-type quarks should be small), while there is less of a
constraint on $M_{14}$ as it has $\Delta I= 0$.

\section{Relevance for short-baseline neutrino oscillation experiments
\label{sec:sbl}}

The present model has mixing only within single families.  In order to explain
the LSND and MiniBooNE electron appearance signals one needs both muon and
electron neutrinos to mix with the same sterile neutrino.  The freedom of
setting a sterile neutrino mass and mixing for one family (the matrix
${\cal S}_2$ in the previous Section) is encouraging for the case of three
families (which must be represented by a $6 \times 6$ matrix).  Furthermore,
if data improve to the extent that three sterile neutrinos are needed to
explain oscillations, \esix~is available.

\section{One neutrino as a possible dark matter candidate \label{sec:dm}}

Another possible use of a third sterile $\nu$ is as a warm dark matter
candidate at the keV scale, as suggested some time ago.\cite{Dodelson:1993je,%
Shi:1998km}  For more recent reviews see Refs.\cite{Abazajian:2012ys,%
Kusenko:2009up}  In contrast to many schemes, the present one distinguishes
between right-hand neutrinos (usually taken very heavy, at the seesaw
scale) and the $n$'s (one of which can easily have keV-scale mass).

There have been two claims for observation of an X-ray line near 3.5 keV.%
\cite{Bulbul:2014sua,Boyarsky:2014jta}  These signals could arise from a
7 keV ``neutrino'' decaying to a photon and a much lighter ``neutrino.''
A corresponding signal is {\it not} seen, however, in the Milky Way.%
\cite{Riemer-Sorensen:2014yda}

There are some special features of \esix~concerning a 7 keV dark matter
candidate.  The Higgs vacuum expectation values considered here correspond to
the five neutral complex scalar bosons in the $27^*$ representation of \esix.
The masses of these bosons are free parameters; two of the five are those of
the minimal supersymmetric standard model or SO(10).  Exchanges of these bosons
can produce the states $n$;  for example, in the processes
$$
d_l + h^c_L \to n_L + N^c_{EL}~~;~~~e^- + E^+_L \to n_L + N^c_{EL}~.
$$
A TeV-scale $Z_N$ produced in the early universe would have appreciable
branching ratio into $n n^c$ pairs, so $n$ are candidates for early
overproduction unless their abundance is diluted by subsequent entropy
production.\cite{Scherrer:1985zt}

\section{Summary \label{sec:sum}}

None of the various hints of sterile neutrinos rises to the level of a
conclusive observation so it is crucial to strengthen or refute them.  Some
effects may be due to interesting non-$\nu$ physics:  for example, if the
MiniBooNE low-energy signal is photons and not electrons.

The grand unified group \esix~[the next step up from SO(10)] naturally
incorporates three candidates for neutrinos with neither left-handed nor
right-handed weak isospin.  \esix~breaking to the standard model times a
particular U(1)$_N$ allows a large Majorana mass for the right-handed neutrino
$N$ and hence the standard seesaw mechanism can proceed without constraints.%
\cite{Ma:1995xk,Callaghan:2013kaa} Masses and mixings of three sterile
neutrinos are at one's disposal to fit oscillation data, assuming present
anomalies are really due to sterile $\nu$ and not something else.

If at most two sterile neutrinos are needed to fit anomalies successfully, a
third is left over as a dark matter candidate.  The \esix~scheme appears to
have enough free parameters to allow such a scenario to successfully
navigate a number of constraints.
 
\section*{Acknowledgments}
I am grateful to H. Fritzsch for the invitation to this conference, and to
Louis Meng and K. K. Phua for generous hospitality.  I thank Janet Conrad,
P. S. Bhupal Dev, Mariana Frank, and Rabi Mohapatra for pointing me to some
helpful literature, and Kevork Abazajian, Joshua Frieman, Richard Hill, Lauren
Hsu, Hitoshi Murayama, and Robert Shrock for useful discussions.  This work was
supported in part by the U. S. Department of energy under Grant No.\
DE--FG02-13ER41598 and in part by funds from the Physics Department of the
University of Chicago.

\end{document}